# DOME Registry: Implementing community-wide recommendations for reporting supervised machine learning in biology

Omar Abdelghani Attafi[1,+], Damiano Clementel[1,+], Konstantinos Kyritsis[2], Emidio Capriotti[3], Gavin Farrell[4], Styliani-Christina Fragkouli[2,5], Leyla Jael Castro[6], András Hatos[7,8,9,10], Tom Lenaerts[11,12,13], Stanislav Mazurenko[14, 15], Soroush Mozaffari[1], Franco Pradelli[1], Patrick Ruch[16,17], Castrense Savojardo[3], Paola Turina[3], Federico Zambelli[18,19], Damiano Piovesan[1], Alexander Miguel Monzon[20], Fotis Psomopoulos[2,*], Silvio C.E. Tosatto[1,21*]

+These authors have equally contributed to this work
*corresponding authors

[1]Department of Biomedical Sciences, University of Padova, Italy,
[2]Institute of Applied Biosciences, Centre for Research and Technology Hellas, Thessaloniki, Greece
[3]Department of Pharmacy and Biotechnology. University of Bologna, Bologna, Italy
[4]ELIXIR Hub, Hinxton, Cambridge, UK
[5]Department of Biology, National and Kapodistrian University of Athens, Athens, Greece
[6]ZB Med Information Centre for Life Sciences, Cologne, Germany
[7]Department of Oncology, Geneva University Hospitals, Geneva, Switzerland
[8]Department of Computational Biology, University of Lausanne, Lausanne, Switzerland
[9]Swiss Institute of Bioinformatics, Lausanne, Switzerland
[10]Swiss Cancer Center Léman, Lausanne, Switzerland
[11]Interuniversity Institute of Bioinformatics in Brussels, Université Libre de Bruxelles-Vrije Universiteit Brussel, Brussels, 1050, Belgium
[12]Machine Learning Group, Université Libre de Bruxelles, Street, Belgium
[13]Artificial Intelligence Laboratory, Vrije Universiteit Brussels, Brussels, Belgium
[14]Loschmidt Laboratories, Department of Experimental Biology and RECETOX, Faculty of Science,
[15]Masaryk University, Brno, Czech Republic International Clinical Research Centre, St. Anne's Hospital, Brno, Czech Republic
[16]HES-SO - HEG Geneva, Geneva, Switzerland
[17]SIB Swiss Institute of Bioinformatics, Geneva, Switzerland
[18]Dept. of Biosciences, University of Milan, Italy
[19]Institute of Biomembranes, Bioenergetics and Molecular Biotechnologies (IBIOM), Bari, Italy.
[20] Department of Information Engineering, University of Padova, Italy
[21] Institute of Biomembranes, Bioenergetics and Molecular Biotechnologies, National Research Council (CNR-IBIOM), Bari, Italy

# Abstract

Supervised machine learning (ML) is used extensively in biology and deserves closer scrutiny. The DOME recommendations aim to enhance the validation and reproducibility of ML research by establishing standards for key aspects such as data handling and processing, optimization, evaluation, and model interpretability. The recommendations help to ensure that key details are reported transparently by providing a structured set of questions. Here, we introduce the DOME Registry, a database that allows scientists to manage and access comprehensive DOME-related information on published ML studies. The registry uses external resources like ORCID, APICURON and the Data Stewardship Wizard to streamline the annotation process and ensure comprehensive documentation. By assigning unique identifiers and DOME scores to publications, the registry fosters a standardized evaluation of ML methods. Future plans include continuing to grow the registry through community curation, improving the DOME score definition and encouraging publishers to adopt DOME standards, promoting transparency and reproducibility of ML in the life sciences. The DOME Registry can be freely accessed from URL: registry.dome-ml.org.

# Introduction

Thanks to the sharp decline in cost for many high-throughput technologies, large volumes of biological data are being generated at a rapid pace and made accessible to researchers. In this context, the field of Machine Learning (ML) or Artificial Intelligence (AI) has risen to prominence given its applicability in data analysis and creation of prediction models using large-scale biological data, such as genomics (1) and proteomics (2) data, thus leading to the development of innovative and far-reaching medical applications (3).

Despite the availability of data and advances in ML/AI, the application of supervised ML algorithms in the biological sciences is still beset by several problems, leading to pitfalls in the wider adoption and reproducibility of these methodologies (4, 5). For example, most ML-related publications are not accompanied by wet-lab experimental validation and are instead based on various approaches of computational assessments (6), which can lead to bias and inaccuracy during result reporting (6). Furthermore, good practices of supervised ML model development, aiming to increase model performance and facilitate generalization and reproducibility, are often overlooked (7, 8). These issues underscore the importance of developing a set of practical recommendations regarding: (i) the construction and evaluation of ML models, considering the utilized data, optimization techniques, and model performance evaluation; and (ii) thorough documentation of the ML development process, encompassing crucial technical details in a comprehensive and concise manner (5).

The ELIXIR Machine Learning Focus Group is part of ELIXIR, the European infrastructure for life science data, which represents over 250 research organizations in 24 countries. Through a community-driven consensus process that involved over 50 ML experts, the ML Focus Group established a set of recommendations for reporting supervised ML approaches in computational biology (9). Collectively known as the DOME recommendations, they cover the major aspects in supervised ML, i.e. data, optimization, model and evaluation, in the context of scientific publications. The DOME recommendations aim to enhance the reproducibility and transparency of published ML approaches for readers, experimentalists, reviewers, and the broader community. Key challenges such as generalization to independent data, effective optimization, and model interpretability are addressed, with an emphasis on rigorous statistical testing for accurate performance assessment. We suggest that DOME recommendations should be used as a report accompanying a manuscript in its supplementary data section, e.g. (10, 11). Currently, the DOME recommendations focus on providing a minimal standard for reporting supervised ML models designed for biological applications and do not extend to other ML classes, e.g. unsupervised or reinforcement learning. A similar effort is the AIMe registry focusing on the description of ML/AI methods for biomedical applications (12). At the time of writing, the AIMe registry contains 35 entries.

Here, we present the DOME Registry to facilitate the dissemination and adoption of DOME recommendations by data scientists and practitioners, working in a wide range of biological fields. The DOME registry constitutes a structured database that allows for researchers to handle and manage DOME-related information on published and unpublished ML applications, providing a unique identifier for each publication and a DOME score. Here, we offer a comprehensive overview of the structure and implementation of the DOME registry, including its connection with external resources such as ORCID identifier (URL: https://orcid.org/), APICURON (13) and the Data Stewardship Wizard (DSW) (17), which enhances user access management and data input. The user-friendly web interface facilitates access to annotations, and supports different user categories for managing, editing, and publishing data. Finally, we present a use case, demonstrating the integration of DOME recommendations and Registry into the article publishing process of a journal.

## Database structure and implementation

### Database structure

The DOME Registry utilizes MongoDB, a NoSQL database, to handle its data. MongoDB's document-oriented structure allows for dynamic and adaptable schema design, which is essential for the evolving nature of the DOME Registry. Figure 1 shows the database schema which is based on two main data structures: annotation and user, and the relationship between these collections. The *annotation* data structure implements the specifications defined in the DOME recommendations (9) for an annotated article. It can be seen as an object containing many properties, which can be split into two different groups. The first group contains descriptive properties for the annotation itself, i.e. unique identifier, short identifier, creation and update timestamps, public flag and the annotation's state. The second group contains the sections defined in the recommendations: dataset, optimization, model and evaluation.

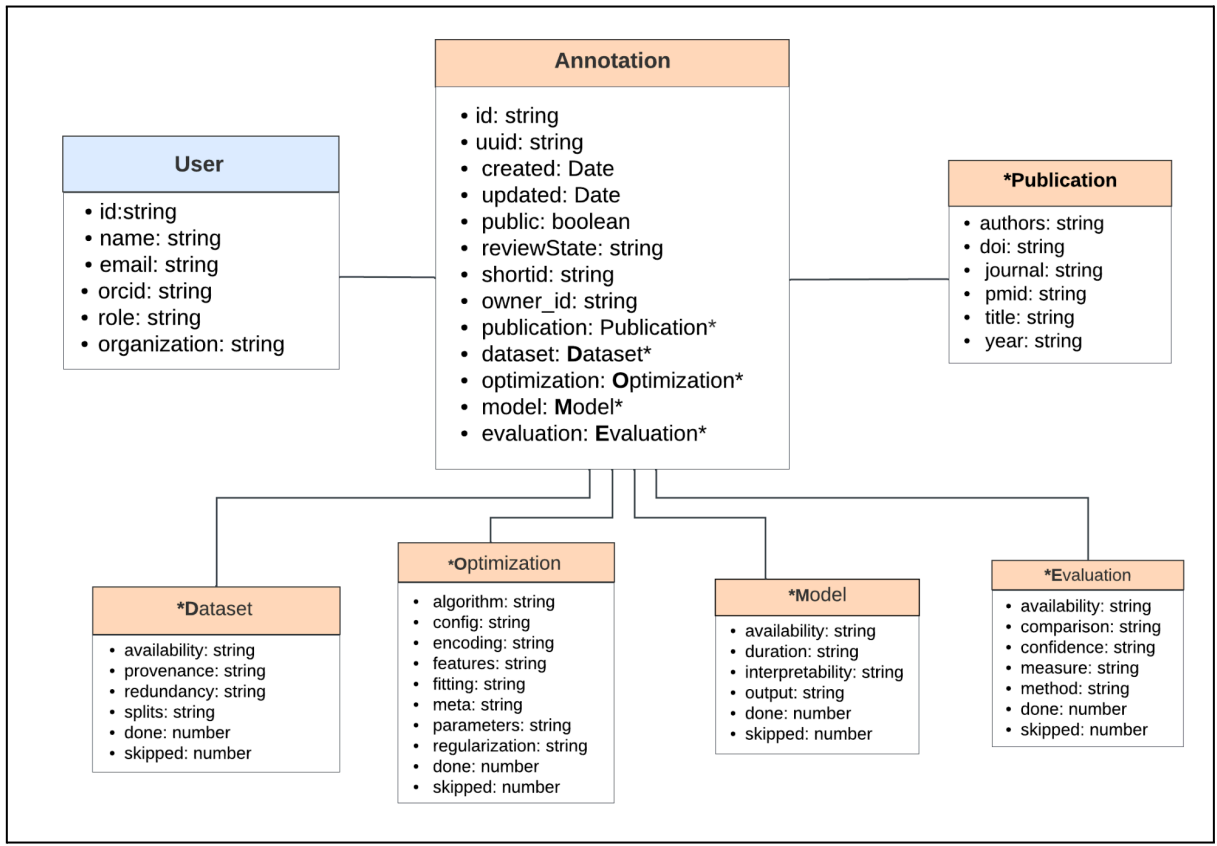

**Figure 1: MongoDB Schema Design for the DOME Registry.** Rectangles of the same color represent fields within the same collection (User and Annotation). The attributes associated with each collection are listed inside the corresponding rectangles. The "owner_id" attribute in the Annotation document references the User document, indicating the creator of the Annotation. For Dataset, Optimization, Model, and Evaluation, there are two additional attributes: "skipped" and "done". The "skipped" attribute tracks the number of fields that are either left empty or marked as "No" or "not assigned." The "done" attribute indicates the number of fields that are correctly filled in. These attributes are used to compute the DOME score.

Each annotation is associated with a user (see Figure 1). The user object is defined by four properties: ORCID identifier, name, email, and organization name (a group of users). The ORCID identifier is uniquely issued by the ORCID authentication service. The association between an annotation and a user is implemented by setting the user's ORCID identifier as a property in the annotation object itself. The system implements three user roles: regular user, user with an organization's admin privileges, and admin. The admin role has access to all annotations in the database (both private and public) and has the authority to delete and modify them regardless of their privacy status. Additionally, the administrator can change the visibility of annotations from private to public. A user with an organization's admin privileges can modify, publish, or delete its own annotations as well as all the annotations of that particular organization. The regular user can only edit over their own annotations and can view only public annotations.

*Web server*

Data in the DOME Registry database is served to the user interface through a web server implementing web APIs compliant with the REST paradigm. Two groups of endpoints have been implemented: one for authentication and authorization and the other for handling annotations.

The authentication and authorization group has two endpoints. The first one redirects to the ORCID authentication service. The second handles the ORCID authentication response, retrieves user information and redirects back to the user interface, filling user data into cookies.

The annotation endpoint group implements CRUD operations (Create, Read, Update, Delete) on annotations stored in the registry (Table 1). Authorized users (e.g. admin) can retrieve, update, delete or insert an annotation by means of the GET, POST, DELETE, and PATCH methods respectively. Users can use all the listed methods. However, the last three methods are available only for the user's own private annotations. The POST method inserts a new annotation into the database, which will be private by default.

The server also executes a series of steps when a new annotation is being inserted into the database. A timestamp and unique identifier are assigned to the new annotation, and the DOME score associated with it is calculated. The DOME score is computed as the number of valid answers to the DOME recommendations, divided by the total number of questions.

The API endpoints page, implemented with SWAGGER UI, describes the DOME registry endpoints to provide a clear visual representation for developers and users. The interface offers an easy-to-use platform for testing and future integration with other services.

| Type | Accessibility | Description |
|------|---------------|-------------|
| Get | Public | Return all the annotations in the database. Parameter shortUID (for a specific annotation) |
| Post | Private | Store a new annotation into the database. |
| Delete | Private | Delete annotation from the database. Parameter uuid (for a specific annotation) |
| Patch | Private | Modify an annotation. Parameter uuid (to modify a specific annotation) |

**Table 1:** DOME registry main endpoints for accessing annotations. All endpoints must begin with the domain fragment https://registry.dome-ml.org/api/.

*User interface*

The DOME Registry is accessed through a user interface (URL:http://registry.dome-ml.org), enabling users to easily search for and retrieve annotations. On the home page, users can view an overview of the number of entries and users who have deposited annotations in the registry. They can also directly access various pages, including statistics, browse, submit, API, about, and help.

The browse page allows users to easily search through annotations. On the top part, a search form allows users to input values, while radio buttons below the text box let them sort results by criteria such as Title, Authors, Year, and DOME Score. The system then searches across all fields in each annotation, displaying only those that match the search terms. Additionally, the statistics page presents various metrics derived from public annotations through interactive plots, including the number of annotated papers per journal, publication year, and DOME score distribution, both overall and by section. For users looking to contribute, the submit page features an introductory video that guides them through creating a new annotation using the DSW and submitting it to the DOME Registry.

Users who have contributed to the DOME Registry by submitting annotations can sign in by clicking the button on the upper right corner of the home page. They will be redirected to the ORCID authentication page, and after a successful login, they will be returned to the DOME Registry home page. The login button will be replaced by the user's name and ORCID identifier, along with a dropdown menu where the logout button can be found. Authenticated users can also choose whether to view public or their own private annotations by using the toggle switch next to the search box on the browse page. Users with admin privileges can access specific annotations through the browse page, where they have the options to edit, delete, and publish directly from the interface.

## DOME Registry workflow

The overall DOME registry system uses various external resources, such as ORCID for researcher identification, Data Stewardship Wizard (DSW) to streamline the process of managing and publishing DOME-related information, and APICURON to provide credit and recognition for biocuration activities. This workflow ensures a structured and validated approach to assessing ML applications in the biological sciences, making it easier for users to manage, edit, and publish relevant data through a user-friendly web interface (see Figure 2).

Users have to login in the DSW to create, share, and submit DOME questionnaires related to a scientific manuscript. The completed questionnaire generates a JSON file containing the relevant user and manuscript data as well as answers for the DOME fields. A validation process is applied to the JSON file through the DOME Registry API, using the ORCID checksum to verify the validity of the ORCID record specified by the user. Following successful validation, the server calculates the DOME score for the annotation, which is saved on the server as private. At this point, the user's contribution is also sent to APICURON for credit attribution.

Admin users can fully manage all annotations received through the DSW by logging into the DOME registry using their ORCID identifiers. They have the ability to review, edit, and delete annotations, and can choose to publish them at their discretion. Furthermore, annotations can be assigned to groups of users within a specific organization (organization admins, e.g. journal publishers) who have admin roles for annotations from their organization, allowing them to determine the publication timing. Only public annotations are accessible to all users through the DOME Registry website.

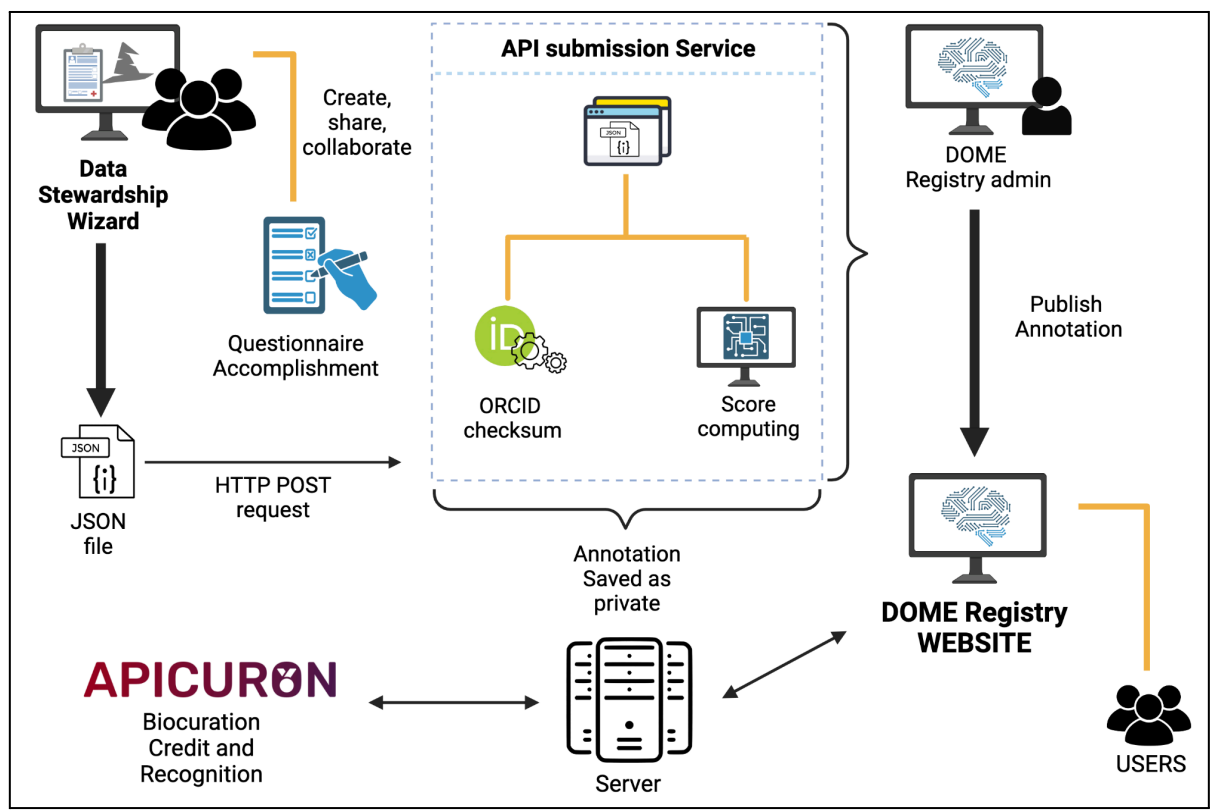

**Figure 2: DOME registry workflow.** Process of creating, validating, and managing DOME annotations with the Integration of ORCID, DSW and APICURON with the DOME Registry API and UI.

*Data Stewardship Wizard*

The Data Stewardship Wizard (DSW) (URL: https://ds-wizard.org/) (14) is a tool designed to facilitate the creation, planning, collaboration, and execution of data management plans (DMPs) based on FAIR principles. The tool aims to simplify the process of building a data management plan, by offering smart questionnaires that guide users through various considerations needed for high-quality research data, without requiring extensive text writing. To facilitate efficient collaborative work, a customized instance of the DSW has been developed specifically for the DOME registry (DOME-DSW), accessible via the URL: https://dome.dsw.elixir-europe.org/. This customized version of DSW enables users to address the set of questions relevant to a particular manuscript in accordance with DOME recommendations. Furthermore, additional information such as the user's ORCID identifier, email address, and manuscript details are required. Users are allowed to create, share, and modify annotations, and upon completion, submit them to the DOME Registry in JSON format (see Figure 1).

*ORCID*

ORCID serves as a component that provides an authentication service and a unique identifier for researchers. It is used within the API Submission Service to verify the identity of contributors and associate them with their published and unpublished works. By integrating ORCID profiles, the system ensures accurate and reliable attribution of user data and annotation contributions, facilitating the management and validation of user-generated content within the DOME Registry.

*APICURON*

APICURON (https://apicuron.org/) (13) is a database to credit and acknowledge the work of biocurators, collecting and aggregating biocuration events from third party resources while generating achievements and displaying leaderboards. APICURON is used as an external tracker of biocuration activities for the DOME registry to increase user engagement and recognize their contributions. The DOME registry is a partner resource of APICURON for the formal recognition of annotation activities.

The only activity recognized in APICURON so far is "annotation submitted", which assigns a score to users who submit annotations from the DSW instance to the DOME registry. Contributions are credited as soon as the annotations are submitted, even if they are still private. APICURON provides two additional recognition components to increase user engagement: medals and badges. Medals are awarded based on relative user rankings, determined by the number of annotations submitted, without fixed quotas. For instance, medals are given to the top annotator, the top 5 annotators, and the top 10 annotators.

Badges, however, are awarded based on fixed thresholds of published annotations, with the progression being 'Newbie Annotator' (2 entries), 'Junior Annotator' (10 entries), 'Senior Annotator' (20 entries), and 'Advanced Contributor' (50 entries).

## Community curation

We adopted a community curation approach to provide the scientific community with an expanded collection of ML publications reported using the DOME recommendations. Specifically, we conducted a targeted query on the Scopus database to retrieve a substantial number of ML-related articles using the following search string: "KEY (machine AND learning) AND KEY (biolog*) AND SUBJAREA (agri OR bioc OR immu OR neur OR phar) AND (LIMIT-TO (OA, "all") OR LIMIT-TO (OA, "BIOC") AND LIMIT-TO DOCTYPE)". Due to the large volume of results (over 4,000 articles), we selected a random subset, divided and distributed among members of the ELIXIR Machine Learning Focus Group. These experts then performed the annotations according to DOME recommendations. By involving multiple experts, the community curation process ensures that the curated data meets high standards of quality and accuracy, leveraging diverse perspectives to minimize bias. Articles that were deemed irrelevant in the context of DOME recommendations, i.e., lacking development and/or application of supervised ML algorithms/models, were excluded from annotation.

## DOME Score

The DOME score attempts to capture the adherence to best practices in the DOME categories, standardizing the evaluation of unsupervised ML research quality and transparency. In its present form, the DOME score is calculated simply as the number of valid answers divided by the total number of questions for each DOME category. The DOME Score should be seen as a preliminary measure rather than a comprehensive assessment. It offers a rough approximation that may not fully capture all the complexities and nuances of ML or specific challenges associated with different biological applications. Future work is needed to continue refining the DOME Score through continuous feedback from the research community, assessing its effectiveness in practical applications, and updating the scoring system to address any identified gaps.

## Use case: Scientific Journal

One direct application of the DOME recommendations and registry is their integration into the journal paper publishing process (i.e. submission, revision, acceptance, publication). A prototypical workflow has been implemented in partnership with GigaScience, which has integrated the DOME Registry in their guidelines for authors regarding ML papers.

When authors submit a manuscript detailing supervised ML approaches, a journal may request the inclusion of the DOME recommendation report. This report can facilitate the revision process, helping the reviewers to better evaluate the merits of the adopted ML methods, and later be provided as supplementary data upon publication. The workflow outlined in Figure 1 and described above can be readily utilized in this context. Submitting authors can log into the DOME-DSW, complete the report indicating it is for a manuscript submission along with the journal name, and then submit it to the registry. Once submitted, the annotation is directly inserted into the DOME Registry database with visibility set as private by default. The reviewers of the manuscript can access the information through a direct link to the private page. Following this, the journal's admin, responsible for reviewing, publishing, deleting, and modifying annotations within their organization, can publish the annotation after review. It is important to note that once the annotation becomes public, the owner rescinds control over it. Subsequently, only the journal's admin retains control over the annotation. A unique identifier is assigned to the annotation when it becomes public in the DOME registry, allowing incorporation as external data source or metadata in the publication.

Integrating the DOME recommendations into the journal publishing process creates a more rigorous, transparent, and reproducible framework for ML research. Journals may benefit from higher-quality submissions and a streamlined review process, reviewers gain a standardized and efficient evaluation tool, and authors receive a structured reporting format that enhances their work's visibility and credibility. This integration not only improves the quality of published ML research but also increases transparency and reproducibility.

## Conclusions and future work

As the field of ML in biological sciences continues to grow, there is an increasing need for standardized reporting practices to ensure that ML is transparent and reproducible. The DOME recommendations aim to enhance the reproducibility and clarity of ML methods, serving as an initial framework for a consensus-based community discussion.

Here we introduced the DOME Registry, which provides a centralized repository for accessing and submitting reports on supervised ML publications. Each entry in the registry includes an annotated DOME recommendations report, along with essential article details such as journal, title, authors, digital object identifier (DOI), and PubMed identifier. Publications are assigned a unique identifier and a DOME score reflecting their adherence to the DOME recommendations. This structured approach not only supports the adoption of DOME recommendations but also addresses the need for reliable evaluation of ML methods. Community curation has further enriched the DOME Registry, resulting in a collection of supervised ML-related publications annotated according to DOME recommendations. This curated collection serves as a reference, promoting best practices in ML model validation and contributing to the broader dissemination of these practices. Future efforts will focus on continuing to expand the registry through community curation, involving researchers from various communities who develop ML methods for life science applications.

Integration with the DSW streamlines the annotation process, allowing users to create, share, and submit annotations in accordance with DOME recommendations. The validation and scoring of annotations ensure their proper completion, and organization admins can manage the publication timing of annotations within their organizations. This process aligns with the goal of engaging publishers to require adherence to DOME recommendations during the article submission process, which will increase the adoption of these standards across different research fields. By providing DOME recommendation reports alongside manuscripts, this approach will enhance the transparency and reproducibility of ML methods from the paper submission stage, through peer review, and up to final publication.

Of course, standards for reporting on Machine Learning models, as well as the tools and services that support these, are of particular significance in the context of FAIR. As the FAIR guiding principles were meant to apply to all digital assets, at a high level, and over time, beyond data they have now been re-interpreted or extended to include the software, tools, algorithms, and workflows that produce data - and recently also adapted in the context of AI models and datasets. Ensuring that data and AI models are FAIR facilitates a better understanding of their content and context, enabling more transparent provenance and reproducibility (15, 16). There is a strong connection between FAIRness and interpretability, as FAIR models facilitate comparisons of benchmark results across models (17) and applications of post-hoc explainable AI methods (18). There are many global efforts towards this direction (19), with key example being the Research Data Alliance Interest Group of FAIR for Machine Learning (URL: https://www.rd-alliance.org/groups/fair-machine-learning-fair4ml-ig), where DOME is one of the standards actively involved, as effective reporting of the ML process is a clear facilitator of the FAIR principles. Finally, it is worth highlighting the necessity of clearly structured and effective metadata for both the FAIR aspects as well as the DOME recommendations. Going beyond the needs of data, additional controlled vocabularies, ontologies and structured metadata are necessary to ensure both well-described ML models, as well as machine-actionable annotation records. There are already a few specific efforts in this direction (20), ultimately allowing the DOME registry to support such metadata.

We believe that the development of standardized reporting guidelines can significantly enhance the transparency, reproducibility and ultimately quality of publications describing ML/AI methods. The DOME Registry represents a step forward in this ongoing endeavor.

## Code Availability

The code underlying the DOME registry is available from its GitHub repository from URL: https://github.com/BioComputingUP/dome-registry/

## Data Availability

The DOME registry and its data can be accessed from URL: registry.dome-ml.org

## Competing Interests

The authors declare that they have no competing interests.

## Acknowledgements

The authors are grateful to GigaScience, in particular Chris Armit, for contributing to the workflow use case of integrating a scientific journal. The ELIXIR Machine Learning Focus Group is acknowledged for insightful discussions as well as anyone who contributed to annotating papers for the DOME registry: Martina Bevilacqua, Piero Fariselli, Pierluigi Martelli, Nils Hoffmann, Rahuman Sheriff Malik Sheriff and Sergio Martinez Cuesta.

## Funding

This work was funded by ELIXIR, the research infrastructure for life-science data. This publication is also partially supported by: COST Action ML4NGP (CA21160), supported by COST (European Cooperation in Science and Technology), European Union's Horizon 2020 research and innovation programme under the Marie Skłodowska-Curie grant agreement No 823886 (REFRACT) and European Union's Horizon Europe research and innovation programme under grant agreement No 101131096 (ELIXIR-STEERS) and under grant agreement No 101129744 (EVERSE). The project is also partially funded by the European Union—NextGenerationEU through 'Italiadomani—PNRR' project CN - G.T.RNA SP. 7 project - National Center for Gene Therapy and Drugs based on RNA Technology (CN3) - [codice MUR CN00000041, C.F. 92315700283], project 'National Centre for HPC, Big Data and Quantum Computing' [codice MUR CN00000013, C.F. C93C22002800006] and Research Infrastructure "ELIXIR x NextGenerationIT: Consolidamento dell'Infrastruttura Italiana per i Dati Omici e la Bioinformatica - ElixirxNextGenIT" - IR0000010 . Tom Lenaerts is supported by Research Foundation - Flanders (FWO) for ELIXIR Belgium (I000323N).